\documentclass[aps,pra,showpacs,reprint,superscriptaddress]{revtex4-1}
\usepackage{graphicx}
\usepackage{bm}
\usepackage{amsmath}

\begin{document}

%Title of paper
\title{Quantum Phase Analysis of Field-Free Molecular Alignment}

\author{Sang Jae Yun}
\email[]{sangjae@kaist.ac.kr}
\affiliation{Department of Physics, Korea Advanced Institute of Science and Technology, Daejeon 305-701, Korea}
\author{Chul Min Kim}
\email[]{chulmin@gist.ac.kr}
\affiliation{Advanced Photonics Research Institute, Gwangju Institute of Science and Technology, Gwangju 500-712, Korea}
\author{Jongmin Lee}
\affiliation{Advanced Photonics Research Institute, Gwangju Institute of Science and Technology, Gwangju 500-712, Korea}
\author{Chang Hee Nam}
%\email[]{chnam@kaist.ac.kr}
\affiliation{Department of Physics, Korea Advanced Institute of Science and Technology, Daejeon 305-701, Korea}

\date{November 5, 2012}

\begin{abstract}
We present quantum mechanical explanations for unresolved phenomena observed in field-free molecular alignment by a femtosecond laser pulse. Quantum phase analysis of molecular rotational states reveals the physical origin of the following phenomena: strong alignment peaks appear periodically, and the temporal shape of each alignment peak changes in an orderly fashion depending on molecular species; the strongest alignment is not achieved at the first peak; the transition between aligned and anti-aligned states is very fast compared to the time scale of rotational dynamics. These features are understood in a unified way analogous to that describing a carrier-envelope-phase-stabilized mode-locked laser. 
\end{abstract}

% insert suggested PACS numbers in braces on next line
\pacs{37.10.Vz, 33.20.Sn, 33.80.--b, 42.50.Hz}
% insert suggested keywords - APS authors don't need to do this
%\keywords{}

%\maketitle must follow title, authors, abstract, \pacs, and \keywords
\maketitle

%\section{Introduction}
Chemical reaction depends on the relative orientation of reactants due to anisotropy in their electronic structure. Likewise, in photochemical processes, the relative orientation between laser polarization and molecular axis is a basic parameter affecting the result. In order to clarify or to control the chemical processes at the most fundamental level, the processes should be examined with aligned molecules. Among several techniques developed so far \cite{Pirani2001, Wu1994, Stapelfeldt2003}, the field-free alignment method using an intense femtosecond laser pulse is now widely adopted. Due to the anisotropy of molecular polarizability, the intense laser pulse induces a torque to rotate molecules to the direction of laser polarization. After interacting with the laser pulse, the molecules periodically align to that direction, achieving field-free alignment \cite{Stapelfeldt2003,Ortigoso1999}. This method has become popular in revealing molecular structures \cite{Itatani2004,Vozzi2011}, controlling and tracing chemical reactions \cite{Larsen1999,Worner2010}, generating high-order harmonics from aligned molecules \cite{Levesque2007,Lee2008}, and compressing optical pulses \cite{Bartels2001,Spanner2003}. 

The field-free alignment method exhibits unique features in the temporal variation of alignment. A typical temporal structure of the method can be seen in Fig.~1, which is obtained by numerically solving the time-dependent Schr\"odinger equation (TDSE) for the rotational state of ${{\rm{O}}_{\rm{2}}}$. The degree of alignment can be represented by $\left\langle {\left\langle {{{\cos }^2}\theta }  \right\rangle } \right\rangle$, in which $\theta $ is the angle between the laser polarization and the molecular axis and the double bracket means that the value is obtained by Boltzmann-averaging of all $\left\langle {{{\cos }^2}\theta } \right\rangle $ (quantum mechanical expectation value) in a thermal ensemble. The unique features can be summarized as follows. First, it shows a revival structure with a full revival time ${T_{rev}}$ in which multiple fractional revivals occur, showing periodic alignment peaks \cite{Seideman1999,Child2003}. Second, each alignment peak has its own temporal profile that changes in an orderly fashion depending on molecular species \cite{Child2003}. Third, the strongest alignment is not achieved at the first peak \cite{Leibscher2003}. Fourth, the transition between aligned and anti-aligned states is very fast compared to the time scale of rotational dynamics \cite{Torres2005}, as marked by $\Delta {T_{tr}}$ in Fig.~1. These features have been observed also in experiments \cite{Bartels2001,Dooley2003}, and many theoretical studies have been made to understand the physical origin of the phenomena \cite{Seideman1999,Child2003,Torres2005,Seideman2001,Renard2004,Renard2005}. Considering that the revival structure is a salient feature of the coherent superposition state of quantum systems, e.g. coherently excited Rydberg atoms \cite{Leichtle1996}, the complex amplitudes of the molecular eigenstates need to be analyzed in detail to explain the features \cite{Seideman2001,Renard2004,Renard2005}.

\begin{figure}[bt]
        \centering\includegraphics[width=8cm]{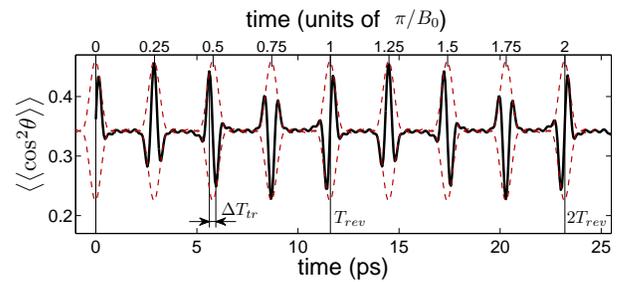}
        \caption{(Color online) Temporal variation of the degree of alignment $\left\langle {\left\langle {{{\cos }^2}\theta }  \right\rangle } \right\rangle$ of ${{\rm{O}}_{\rm{2}}}$ molecules at 90 K after interacting with a ${\sin ^2}$-type laser pulse with a duration of $30 ~ {\rm{ fs}}$ (FWHM) and a cycle-averaged peak intensity of ${\rm{5 \times 1}}{{\rm{0}}^{{\rm{13}}}}$ ${\rm{W/c}}{{\rm{m}}^{\rm{2}}}$.}
\end{figure}

In this work we present an analysis of the molecular alignment induced by a femtosecond laser pulse in terms of the quantum phase of molecular rotational states. By numerically solving the TDSE of molecular rotational states, the quantum phase of rotational eigenstates is obtained. With the phase of the eigenstates known, the temporal variation of molecular alignment is described in a way similar to that of mode-locked laser pulses. The results provide the quantum mechanical explanation on all the above features in a unified way, signifying that the quantum phase of rotational states plays a crucial role in understanding the alignment dynamics.

To be specific, the alignment of an ${{\rm{O}}_{\rm{2}}}$ molecule, a typical non-polar linear molecule, by a femtosecond laser pulse was considered for analysis. The TDSE for such a case within a rigid rotor model in atomic units is given by \cite{Stapelfeldt2003,Lemeshko2010}
\begin{equation}
i{d \over {dt}}\left| {\Psi (t)} \right\rangle  = {B_0}\left( {{{\bf{J}}^2} - { I \over I_0 }g(t){{\cos }^2}\theta } \right)\left| {\Psi (t)} \right\rangle ,
\end{equation}
where $B_0$ is the molecular rotational constant, $I$ the cycle-averaged peak intensity, $I_0 \equiv {B_0 c} /\left( 2\pi \Delta \alpha \right)$ the natural intensity scale of the rotational molecular Hamiltonian, and $g(t) = {\sin ^2}\left( {\pi t/\tau } \right)$ the normalized laser-intensity profile. $\Delta \alpha ( \equiv {\alpha _\parallel } - {\alpha _ \bot })$ is the anisotropy of molecular polarizability. ${{\rm{O}}_{\rm{2}}}$ molecule has ${B_0} = 2.856 \times {10^{ - 16}}{~\rm{ erg ~(}}1.437{~\rm{c}}{{\rm{m}}^{ - 1}}{\rm{)}}$ and  $\Delta \alpha  = 1.12 ~ {\rm{ }}{{\rm{{\AA}}}^3}$ \cite{NIST}, giving a natural rotational time scale $\pi /{B_0} = 11.6~{\rm{ps}}$ and $I_0=1.22 \times 10^{11}{~\rm{W/cm^2}}$. In our calculation, the condition of $\tau=60~{\rm{fs}} = 0.0052  \pi /{B_0}$ and $I = 5 \times {10^{13}}{~\rm{ W/c}}{{\rm{m}}^2} = 410  I_0$ was used.

The evolution of molecular rotational states during the laser interaction is investigated by considering an angular momentum eigenstate $\left| {{J_i},{M_i}} \right\rangle $ as an initial state. Since the laser interaction causes sequential Raman-type transitions with the selection rules $\Delta J = 0, \pm 2$ and $\Delta M = 0$ \cite{Stapelfeldt2003}, the rotational state under interaction can be written as
\begin{equation}
\left| {\Psi (t)} \right\rangle  = \sum\limits_{n = {n_0}}^\infty  {\sqrt {{P_{{J_i} + 2n}}(t)} {e^{i{\Phi _{{J_i} + 2n}}(t)}}\left| {{J_i} + 2n,{M_i}} \right\rangle } ,
\end{equation}
where ${P_{{J_i} + 2n}}$ and ${\Phi _{{J_i} + 2n}}$ are the probability and the quantum phase of $n$-th emerging component, respectively. The emerging components with positive $n$ come from Stokes processes,
while those with negative $n$ from anti-Stokes processes. Since ${J_i} + 2n$ cannot be smaller than $\left| {{M_i}} \right|$, the anti-Stokes processes are bound by a non-positive integer ${n_0}$ \cite{Lemeshko2010}. Inserting Eq.~(2) into Eq.~(1) gives coupled differential equations of ${P_{{J_i} + 2n}(t)}$ and ${\Phi _{{J_i} + 2n}(t)}$. In obtaining the numerical solution, the Crank-Nicolson method \cite{Press1992} was used. Figure 2 shows the numerical solution of the rotational state for an initial state $\left| {3,0} \right\rangle $ as an example.

\begin{figure}[bt]
        \centering\includegraphics[width=8cm]{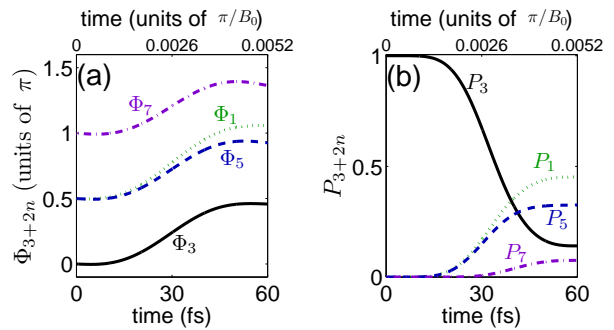}
        \caption{(Color online) Temporal evolution of the rotational state initially at $\left| {{J_i} = 3,{M_i} = 0} \right\rangle $ of ${{\rm{O}}_{\rm{2}}}$ during laser interaction: (a) phase and (b) probability. The laser condition is the same as that in Fig.~1. Time 0 and 60 fs correspond to the beginning and the end of the laser pulse, respectively.}
\end{figure}

The quantum phase ${\Phi _{{J_i} + 2n}}$ is the basic element of the alignment dynamics. As seen in Fig.~2~(a), the quantum phase of $n$-th emerging component starts with $\pi \left| n \right|/2$ \cite{Seideman2001,Renard2004,Renard2005}. This phenomenon can be understood by analyzing the mathematical form of Eq.~(1). The emerging components are generated through the off-diagonal matrix elements of the interaction Hamiltonian, $-B_0 I g(t){{\cos }^2}\theta / I_0 $, which are always negative as the matrix elements of ${\cos ^2}\theta $ are positive \cite{Owschimikow2011}, and $\Delta \alpha $ is also positive for most linear molecules including ${{\rm{O}}_{\rm{2}}}$. Because of this negativity of the interaction Hamiltonian and the factor $i = \exp (i\pi /2)$ in the left-hand side of Eq.~(1), the starting phase of an emerging component must add ${\pi/2}$ to the phase of its parent component. Since a new component arises only through sequential transitions, ${\pi/2}$ is accumulated to its phase whenever a succeeding component emerges. Even when the populating process is of anti-Stokes $\left( {n < 0} \right)$, the phase accumulation is also ${\pi/2}$, not $-{\pi/2}$, because the matrix element of the interaction Hamiltonian is equally negative. Thus, the $n$-th quantum phase starts at the value of $\pi \left| n \right|/2$.

Since the alignment is an interference among the rotational eigenstates, not the quantum phase itself but the relative phase determines the dynamics. As the laser interaction goes on, the quantum phase itself varies significantly, but the phase difference of ${\pi/2}$ between neighboring components set at the beginning does not change much because the phase-varying rates are comparable to each other, and the laser duration is very short compared to the time scale of the rotational dynamics. Thus, until the end of the laser pulse, the phase difference between neighboring components is kept to be around ${\pi/2}$ \cite{Seideman2001,Renard2004,Renard2005}, as observed in Fig.~2~(a). 

After the interaction, the molecular wavefunction evolves freely and the dynamics of alignment is determined solely by the molecular state at the end of the interaction (t = 60 fs in Fig.~2). Setting this moment as the new time origin, the field-free evolution of the degree of alignment, i.e. expectation value of ${{{\cos }^2}\theta }$, is given by
\begin{gather}
\left\langle {{{\cos }^2}\theta } \right\rangle (t) = \sum\limits_{n = {n_0}}^\infty  {P_{{J_i} + 2n}^{}C_{{J_i} + 2n,{J_i} + 2n}^{}} \nonumber
\\
+ \left\{ {\sum\limits_{n = {n_0}}^\infty  {{1 \over 2}\sqrt {I_n^{}} } {e^{i\left( {\omega _n^{}t - \varphi _n^{}} \right)}} + {\rm{c}}.{\rm{c}}.} \right\},
\end{gather}
where
\begin{gather}
I_n^{} = 4P_{{J_i} + 2n}^{}P_{{J_i} + 2n + 2}^{}{\left( {C_{{J_i} + 2n,{J_i} + 2n + 2}^{}} \right)^2},
\\
\omega _n^{} = {E_{{J_i} + 2n + 2}} - {E_{{J_i} + 2n}} = {B_0}(8n + 4{J_i} + 6),
\\
\varphi _n^{} = \Phi _{{J_i} + 2n + 2}^{} - \Phi _{{J_i} + 2n}^{},
\end{gather}
$C_{{J_a},{J_b}}^{} \equiv \left\langle {{J_a},{M_i}} \right|{\cos ^2}\theta \left| {{J_b},{M_i}} \right\rangle $, and ${E_J} \equiv {B_0}J(J + 1)$. Here, $P_{{J_i} + 2n}$ and $\Phi _{{J_i} + 2n}$ are the values evaluated at the end of the laser pulse. Because Eq.~(3) is expressed in a complex Fourier series, the time-domain behaviour can be analyzed in the frequency-domain, where the spectral intensity $I_n$, frequency $\omega_n$, and spectral phase $\varphi_n$ are given by Eqs.~(4), (5), and (6), respectively. The first term of Eq.~(3) is the DC term giving an offset in the time-domain, whereas the second term is the oscillating part having multiple frequencies. Equation (5) expresses an equally spaced frequency comb with a comb spacing of $8{B_0}$. 

\begin{figure}[tb]
        \centering\includegraphics[width=8cm]{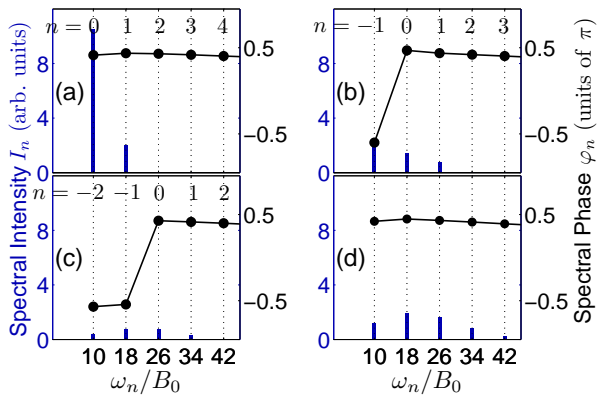}
        \caption{(Color online) Spectral intensity (bars) and phase (filled circles) of $\left\langle {{{\cos }^2}\theta } \right\rangle (t)$ generated from several initial states with ${M_i}=0$ after the laser interaction with molecules at 90 K: (a) ${J_i} = 1$, ${M_i} = 0$; (b) ${J_i} = 3$, ${M_i} = 0$; (c) ${J_i} = 5$, ${M_i} = 0$; and (d) Boltzmann-averaged spectrum from all the ${J_i}$ states with ${M_i} = 0$ at 90 K. The laser condition is the same as that in Fig.~1. The illustrated spectral intensity ${I_n}$ was obtained by multiplying the corresponding Boltzmann weighting factor to Eq.~(4).}
\end{figure}

The spectral phase ${\varphi _{n}}$ of $\left\langle {{{\cos }^2}\theta } \right\rangle (t)$ in Eq.~(3) is the relative phase of two neighboring quantum states, as expressed in Eq.~(6). ${\varphi _{n}}$ has nearly binary values close to either ${\pi/2}$ or $ - {\pi/2}$ \cite{Renard2004,Renard2005}, as shown in Figs.~3~(a)-(c). This is explained by the initial quantum phases $\pi \left| n \right|/2$ of which relative difference is mostly maintained until the end of the laser pulse. Because of this, Stokes processes $\left( {n \ge 0} \right)$ result in ${\varphi _{n}}$ near ${\pi/2}$ whereas anti-Stokes processes $\left( {n < 0} \right)$ generate near $ - {\pi/2}$, making them nearly out of phase. This is the most critical part in understanding the alignment dynamics. 

In order to obtain experimentally observable alignment dynamics, $\left\langle {{{\cos }^2}\theta } \right\rangle (t)$ should be averaged over an initial thermal ensemble to give $\left\langle {\left\langle {{{\cos }^2}\theta }  \right\rangle } \right\rangle (t)$ as shown in Fig.~1. Instead of considering all the initial states at once, it is more systematic to consider only part of them by dividing into groups according to ${M_i}$. Figure 3 illustrates a group with ${M_i} = 0$. For a given ${M_i}$, the lowest-${J_i}$ state generates only Stokes processes so that the spectral phase is nearly constant, ${\pi/2}$, as shown in Fig.~3~(a). Anti-Stokes processes produce low-frequency components and can happen only from initial high-${J_i}$ states of which Boltzmann factors are smaller than that of the lowest-${J_i}$ state. Because of this, after cancelling out the spectral intensity of low-frequency parts due to the opposite phase between Stokes and anti-Stokes processes, the resulting ensemble-averaged phase follows the phase of the Stokes processes as can be seen in Fig.~3~(d). In contrast to the low-frequency parts, the spectral intensity of high-frequency parts add up constructively because only Stokes processes prevail in that region. Consequently, the ensemble-averaged phase of Fig.~3~(d) is formed almost identical to that of Fig.~3~(a). Similarly, other groups having different ${M_i}$ also show the same behavior, nearly constant phase of ${\pi/2}$. Owing to this phase coherence, the resulting spectral intensity from all the initial states is effectively enhanced. This is the main reason for achieving the strong alignment. The resulting spectra of $\left\langle {\left\langle {{{\cos }^2}\theta }  \right\rangle } \right\rangle (t)$ from all the initial states are shown in Fig.~4, where almost flat spectral phase and equally spaced frequency comb are observed. By synthesizing the spectral components in Fig.~4, the complex temporal variation in Fig.~1 was obtained.

\begin{figure}[tb]
        \centering\includegraphics[width=9cm]{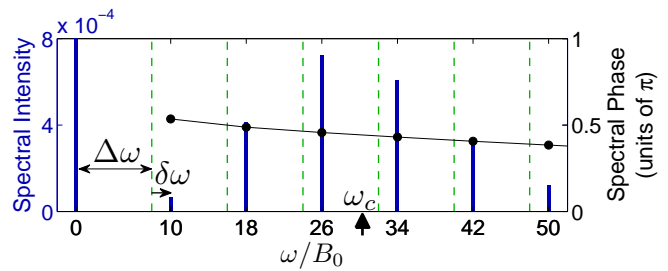}
        \caption{(Color online) Spectral intensity and phase of Boltzmann-averaged $\left\langle {\left\langle {{{\cos }^2}\theta }  \right\rangle } \right\rangle (t)$ from all initial states at 90~K. The laser condition is the same as that in Fig.~1. } 
\end{figure}

The complex temporal dynamics of molecular alignment after the laser pulse can be understood by treating it as a mode-locked signal with almost flat phase. Since an equally spaced frequency comb having a linear spectral phase produces a pulse train in the time-domain \cite{Weiner2009}, it is natural that $\left\langle {\left\langle {{{\cos }^2}\theta }  \right\rangle } \right\rangle (t)$ in Fig.~1 also shows pulsed shape like mode-locked laser pulses, exhibiting periodic strong alignment peaks. 

The temporal profile of each alignment peak in Fig.~1 can be analyzed by using the analogy to the carrier-envelope phase (CEP) of a mode-locked laser pulse. In Fig.~1, the solid line corresponds to the carrier oscillation, while the dashed line the envelope. It is well known that the CEP-sequence is completely determined by the frequency-comb structure \cite{Jones2000,Weiner2009}. Let us denote the frequency-comb as ${\omega _m} = m\Delta \omega  + \delta \omega $ ($m = $
1, 2,$ \cdots $), where $\Delta \omega $ is the pulse-repetition rate (multiplied by $2\pi $), and $\delta \omega $ is the offset frequency as shown in Fig.~4. Then, in the time-domain, the increment of CEP from one pulse to the next is $\Delta {\phi _{CEP}} =  - 2\pi  \cdot \delta \omega /\Delta \omega $ \cite{Weiner2009}. Since the full revival occurs when the accumulated CEP change reaches $2\pi $, the number of fractional revivals occurring during a full revival period should be $N = 2\pi /\left| {\Delta {\phi _{CEP}}} \right| = \Delta \omega /\left| {\delta \omega } \right|$, and the full revival time should be ${T_{rev}} = 2\pi N/\Delta \omega  = 2\pi /\left| {\delta \omega } \right|$. The comb parameters, $\Delta \omega $ and $\delta \omega $, are obtained from Eq.~(5) for a given molecular species. Since an ${{\rm{O}}_{\rm{2}}}$ molecule has only odd-$J$ states due to its nuclear spin statistics \cite{Herzberg1989}, $\Delta \omega  = 8{B_0}$ and $\delta \omega  = 2{B_0}$ as shown in Fig.~4, which results in $\Delta {\phi _{CEP}} =  - {\pi/2}$, $N = 4$, and ${T_{rev}} = {\pi / {{B_0}}}$. 

The CEP of the first alignment peak is always nearly ${\pi/2}$ (sine-like) regardless of molecular species, and it relates to the fact that the strongest alignment is not achieved at the first peak. The first CEP is determined by the temporal phases of all the frequency components at $t=0$. These phases at $t=0$ coincide with the spectral phases as can be deduced from Eq.~(3). Since all the spectral phases lie near ${\pi/2}$, the first CEP should also be near ${\pi/2}$. With the results of the last paragraph, the CEP sequence of ${{\rm{O}}_{\rm{2}}}$ molecule should be $\left( {{\pi/2},0, - {\pi/2}, - \pi } \right)$. This explains why the strongest alignment does not occur at the first peak because it occurs when the CEP is $0$. The same consideration gives the CEP sequence $\left( {{\pi/2},\pi , - {\pi/2},0} \right)$ for even-$J$ molecules such as ${\rm{C}}{{\rm{O}}_{\rm{2}}}$ \cite{Bartels2001} and $\left( {{\pi/2}, - {\pi/2}} \right)$ for all-$J$ molecules such as ${{\rm{N}}_2}$ \cite{Dooley2003}.

Our analysis also provides a detailed explanation on temperature dependence of molecular alignment. It seems intuitively obvious that the degree of alignment decreases with temperature. However, although this phenomenon was partially explained from a classical treatment \cite{Leibscher2004}, no quantum mechanical explanation has been given. Here it can be explained by noting that Stokes and anti-Stokes processes generate opposite spectral phases. At a high temperature, considerable population is at high-${J_i}$ states initially. A higher-${J_i}$ state causes more anti-Stokes processes that diminish the spectral intensity of low-frequency part. The higher the temperature is, the wider the frequency range is diminished. Because a small spectral intensity is equivalent to a low oscillating amplitude in the time-domain, the diminished spectral intensity results in weaker alignment.  

The fast transition between aligned and anti-aligned states can be explained by considering the center frequency of the frequency comb, ${\omega _c}$ in Fig.~4. ${\omega _c}$ corresponds to the carrier oscillation period with the relation $\Delta {T_{tr}} = {\pi  \mathord{\left/
 {\vphantom {\pi  {{\omega _c}}}} \right.
 \kern-\nulldelimiterspace} {{\omega _c}}}$. It means that high ${\omega _c}$ is equivalent to fast transition. High ${\omega _c}$ can result from either a high temperature or a strong laser interaction. A high temperature results in a high ${\omega _c}$ with two reasons. First, initially existing high-${J_i}$ states easily generate high-frequency parts because, from Eq.~(5), ${\omega _{n}}$ becomes a high frequency with even a small $n$. Second, the spectral intensity of low-frequency part is diminished as explained before. A strong laser field also causes the high ${\omega _c}$ because the strong interaction populates much higher-$J$ states through high-$n$ Stokes transition so that the spectral intensity in the high-frequency region increases. Thus, $\Delta {T_{tr}}$ becomes shorter when the temperature is high or the laser interaction is strong. In Fig.~4, ${\omega _c}$ is $29.8{B_0}$ leading to  $\Delta {T_{tr}} = 388~{\rm{ fs}}=0.034 T_{\rm{rev}}$. This fast transition is not surprising because the superposition of the eigenstates of $J = 7$ and $J = 9$ generates ${\omega _c} = 34{B_0}$ leading to a transition time of $341~{\rm{ fs}}~ (=0.030 T_{\rm{rev}})$. 

The molecular alignment at higher intensity was also examined. For example, at $I = 8 \times {10^{13}}~{\rm{ W/c}}{{\rm{m}}^2} = 655 I_0$, the quantum phase ${\Phi _{{J_i}}}$ showed an abrupt jump due to the redistribution of rotational states during the laser interaction. Due to the phase jump, the ${\pi/2}$ phase-difference between adjacent quantum phases did not hold any more, and the spectral phase of $\left\langle {{{\cos }^2}\theta } \right\rangle (t)$ deviated from that of Figs.~3~(a)-(c). It was, however, found that the Boltzmann-averaging mitigate these effects to make the ensemble-averaged spectral phase to be almost the same as that of Fig.~3~(d). 

In conclusion, we have presented clear quantum mechanical explanations on the puzzling phenomena observed in the field-free molecular alignment. Analyzing the quantum phase of molecular rotational states, we have shown that $\left\langle {\left\langle {{{\cos }^2}\theta }  \right\rangle } \right\rangle (t)$ has almost flat spectral phase and equally spaced frequency comb. All the features of complex alignment dynamics could be explained in an analogous way describing a CEP-stabilized mode-locked laser. The analysis given here can be extended further to more challenging subject such as orientation of polar molecules or three-dimensional alignment of asymmetric top molecules. The clear understanding for alignment dynamics will bring the optimization of experimental conditions and the development of a better method controlling molecular rotational states. 

\begin{acknowledgments}
This work was supported by the Ministry of Education, Science and Technology of Korea through the National Research Foundation and by the Ministry of Knowledge and Economy of Korea through the Ultrashort Quantum Beam Facility Program.
\end{acknowledgments}

\bibliography{MR4}

%\begin{thebibliography}{24}%
%\end{thebibliography}%

\end{document}